# Inducing transparency in the films of highly scattering particles

Talha Erdem[1,*], Lan Yang[2], Peicheng Xu[1], Yemliha Altintas[3], Thomas O'Neil[1], Alessio Caciagli[1], Caterina Ducati[4], Evren Mutlugun[3], Oren A. Scherman[2], Erika Eiser[1,*]

[1]*Cavendish Laboratory, Department of Physics, University of Cambridge, JJ Thomson Avenue, Cambridge CB3 0HE, United Kingdom*

[2]*Melville Laboratory for Polymer Synthesis, Department of Chemistry, University of Cambridge, Lensfield Road, Cambridge CB2 1EW, United Kingdom*

[3]*Abdullah Gül University, Departments of Electrical-Electronics Engineering, and Materials Science and Nanotechnology Engineering, 38080, Kayseri, Turkey*

[4]*Department of Material Science and Metallurgy, University of Cambridge, 27 Charles Babbage Road, Cambridge CB3 0FS, United Kingdom*

*Corresponding Author: te294@cam.ac.uk and ee247@cam.ac.uk

Today colloids are employed in various products from creams and coatings to electronics. The ability to control their chemical, optical, or electronic features by controlling their size and shape explains why these materials are so widely employed. Nevertheless, altering some of these properties may also lead to some undesired side effects, one of which is an increase in optical scattering upon concentration. Here, we address this strong scattering issue in films made of colloids with high surface roughness. We focus on raspberry type polymeric particles made of a spherical polystyrene core decorated by small hemispherical domains of acrylate. Owing to their surface charge and model 'roughness', aqueous dispersions of these particles display an unusual stability against aggregation. Under certain angles, their solid films display a brilliant red color due to Bragg scattering but otherwise appear completely white on account of strong scattering. To suppress the scattering and induce transparency, we prepared films by hybridizing them either with oppositely-charged PS-particles that fit the length-scale of the raspberry roughness or with quantum dots. We report that the smaller PS-particles prevent raspberry particle aggregation in solid films and suppress scattering by decreasing the spatial variation of the refractive index. We believe that the results presented here provide a simple strategy to suppress strong scattering of rough particles and allow for their utilization in optical coatings, cosmetics, or photonics.

**Introduction**

Colloidal nanoparticles are commonly employed in various applications including optical coatings,[1] optoelectronic devices,[2-3] sensing,[4-7] photovoltaics,[8-10] electronics,[11-13] and bio-imaging,[14-18] to name a few. Today, they possess an invaluable place in various branches of science and engineering as they enable control over reactivity, stability, size, and shape, which allows for tuning their physical, electronic, chemical, and optical features.

An important advantage colloidal nanoparticles offer is the increased surface area per volume, i.e. surface-to-volume ratio. Increasing this ratio enables improved stability in dispersions,[19] enhances interaction with surrounding materials,[20] and helps transfer charges[21] that may be especially of interest for sensing and photovoltaic applications. One of the strategies to increase the surface-to-volume ratio is decreasing the size of the nanoparticles as this ratio is proportional to the inverse of the radius for a spherical particle. Nevertheless, this approach may be sometimes unpractical or undesirable. Another strategy that is commonly exploited to increase this ratio is utilizing nanoparticles with rough surfaces. Adding bumps onto the surfaces of smooth particles boosts the surface-to-volume ratio, which offers advantages without substantially changing the size of the colloids. For example, we previously showed that aqueous suspensions of raspberry particles (called RBs hereafter) that are made of a spherical polystyrene core-particle surrounded by many smaller polyacrylate nanoparticles exhibit unexpected stability even at significantly high salt concentrations.[22] This improvement was enabled by the high surface charge of these particles along with their significantly improved surface-to-volume ratio.

Increasing the surface roughness, however, comes at the expense of increasing the scattering of the RBs, which may not be a desired feature for some applications. To address this problem, we present here a simple methodology to suppress the optical scattering of the solid films by hybridizing these raspberry particles with either polymeric nanoparticles or quantum dots. Our strategy relies on embedding the RBs into a matrix of smaller, oppositely-charged nanoparticles whose solid films possess a transparent character. We show that when these nanoparticles are bound to our RB-particles via Coulomb attractions - the resulting composite raspberry particles do not aggregate in a regular array but suppress the strong scattering observed in pure RB-particle films. Our experiments, supported by electromagnetic simulations, showed that placing RBs into a matrix of smaller nanoparticles significantly limits the scattering angle of the raspberry particles and thus induces transparency in their solid films. We believe that the simple methodology we propose here may make various types of organic and inorganic nanoparticles suitable for photonic and electronic devices by suppressing their scattering.

**Experimental Methods**

*Synthesis of raspberry particles.* Following the synthesis introduced by Lan et al.[22] 2.08 g of styrene (St, 2.08 g, 20.0 mmol), 130 mg of divinylbenzene (DVB) and 428 mg 2-(methacryloyloxy) ethyl acetoacetate (AM) were loaded into a 50 mL water/ethanol mixture with a ratio of 80/20 v/v%. Subsequently, 54 mg of 2,2'-azobis(2-methylpropionamide) dihydrochloride (AIBA) was added to the mixture, kept under $N_2$ atmosphere for 1 h, and then heated to 70 °C. The reaction continued for 24 h. Synthesized raspberry particles were then purified by dialysis against water for 5 d.

*Synthesis of negatively charged polystyrene nanoparticles.* Sodium 4-vinylbenzenesulfonate (SVBS, 660 mg) and potassium persulfate (KPS, 135 mg) were dissolved in 100 mL deionized water. The mixture was then sealed and degassed by cycling vacuum and nitrogen five times, before being heated to 70 °C with vigorous stirring. Styrene (5.2 g) was injected immediately. After 24 h, the reaction was stopped by quenching the reaction vessel on ice and the nanoparticles were washed by dialysis.

*Synthesis of positively charged polystyrene nanoparticles.* Methyl iodide (5ml) dissolved in dichloromethane (DCM, 50 mL) was added dropwise to a solution of bipyridine (10 g) in 100 mL DCM with vigorous stirring and left to react for 24 h. 1,1'-dimethyl-[4,4'-bipyridine]-1,1'-diium iodide (MV+) was collected by filtration and drying. A solution of MV+ (6.02 g) in acetonitrile (300 mL) was prepared at 80 °C and allowed to cool to 60 °C. 4-vinylbenzyl chloride (10ml) was added and the reaction allowed to proceed for 24 h. 1-methyl-1'-(4-vinylbenzyl)-[4,4'-bipyridine]-1,1'-diium chloride iodide (StMV) was collected by filtration and drying.

StMV (7 mg for 280 nm particles, 90 mg for 45 nm particles) and 2,2'-Azobis(2-methylpropionamidine) dihydrochloride (AIBA) (25 mg) were dissolved in 100 mL water and the solution degassed by cycling vacuum and nitrogen five times. The solution was heated to 80 °C with vigorous stirring and styrene (5.2 g) was injected immediately. After 24 h, the reaction was stopped by quenching the reaction vessel on ice and the nanoparticles were washed by dialysis.

*Synthesis of InP/ZnS core/shell nanocrystal quantum dots.* InP/ZnS quantum dots were synthesized using a recipe proposed by Altıntas et al.[23] with some modification of the synthesis procedure. 0.12 mmol of indium acetate, 0.36 mmol of myristic acid and 6 mL of 1-octadecene (ODE) were mixed in a three necked 25 mL flask, degassed for 5 min, and then purged with Ar-gas for 5 min. This procedure was repeated three times to remove dissolved oxygen and other impurities. Subsequently, the solution was heated to 100 °C under vacuum ($3\times10^{-3}$ mbar) and kept at this pressure for 1.5 h to obtain a clear solution. This clear solution was then cooled down to room temperature. A successive vacuum-gas procedure was used and the temperature was increased to 220 °C after addition of 1 mmol of Zn-stearate purum and 0.025 mmol of 1-dodecanethiol (DDT). Tris-(trimethylsilyl) phosphine (($TMS)_3P$) as a phosphor source (0.08 mmol of $(TMS)_3P$ in 1 mL of ODE prepared in the glovebox) was quickly injected into the flask after reheating to 220 °C. Next, the InP core quantum dots were grown at 285 °C for 10 min. After cooling down the flask to room temperature, the shell was coated by adding 0.2 mmol of Zn-stearate (purum) at room temperature and then applying a successive vacuum-gas conversion procedure. The temperature of the flask was increased to 230 °C under Ar atmosphere and kept for 3 h. 0.4 mmol of DDT was added dropwise (prepared in 1 ml ODE) into the flask and kept further for 1 h.

After cooling back to room temperature, the crude solution of the synthesis was transferred into a 50 mL centrifuge tube by adding 5 mL of hexane and centrifuged two times at 5000 rpm for 8 min to separate unreacted species via discarding the precipitated particles. 25 mL of acetone and 3 mL of methanol was added into the supernatant solution and then precipitated by centrifugation at 5000 rpm for 10 min and dissolved in hexane. Precipitation method was carried out two times with the method described and finally the thus synthesized QDs were re-dispersed in hexane.

*Phase transfer of the quantum dots.* The QDs dispersed in chloroform were transferred to the water phase using a methodology described in ref. [24]. The procedure involved coating the QDs with an amphiphilic polymer. This polymer was synthesized by first dissolving 3.084 g of poly(isobutylene-alt-maleic anhydride) and 3.5 mL dodecylamine in 100 mL of tetrahydrofuran (THF) in a round flask. The mixture was vigorously stirred and then kept at 60 °C for 3 h; subsequently the reaction mixture was concentrated to approximately five times its original QD content using a rotary evaporator and kept at 60 °C overnight. Subsequently, the mixture was completely dried using again a rotary evaporator and dissolved in 25 mL of chloroform such that the final concentration of the monomer units becomes 0.8 M. Next, 250 µL of the QDs (~20 µM) was mixed with 0.8 mL of the synthesized polymer and the mixture was slowly evaporated using a rotary evaporator. The resulting solid was then dissolved in 0.1 M NaOH and sonicated. To remove large aggregates, the mixture was filtered using a 0.22 µm syringe filter. Finally, the QDs were re-dispersed into a TAE buffer by using an ultracentrifuge filter unit (100 kDa).

*Molar concentration estimates of the nanoparticles.* The molar concentration of the RBs was estimated by calculating their molecular weight and measuring the density of the dispersion. For this purpose, we first calculated the number of polyacrylate half spheres of 45 nm diameter that can fit on the surface of 235 nm sized PS nanoparticles. The number was obtained by a Monte Carlo simulation of hard disks near random close packing using spherical boundary conditions, according to the scheme described in ref. [25]. The radius of the PS nanoparticle $R_{PS}$ and number of hard disks N were fixed at the beginning of the simulation and the scheme allowed for displacement and density moves, in which the half spheres' radius $R_{HS}$ was increased until equilibrium was reached. We obtained the optimal number of hard spheres $N_{op}$ as the value at which the ratio $|R_{HS}/R_{PS}|$ at equilibrium matched the experimental value. In our case, $N_{op} = 90$. Subsequently, we calculated the molecular weight of a raspberry particle by taking the density of the polyacrylate to be 1.2 g/cm³ and the density of polystyrene to 1.05 g/cm³. We then measured the density of our 3 wt% raspberry nanoparticle stock solution and used this information to determine the concentration as 3 nmol/L. The molar concentrations of the positively and negatively-charged PS nanoparticles were estimated by calculating the molecular weight of the 45 nm large PS particles the and 280 nm RBs using the density of the polymer as 1.05 g/cm³. We next measured the density of the 5 wt% nanoparticle solution and estimated the molar concentration to be 500 nM. The concentration of the InP/ZnS quantum dots was estimated using the absorption cross section given in ref. [26].

*Formation of solid films.* Prior to film formation, nanoparticles (negatively-charged PS particles, positively-charged PS nanoparticles, or InP/ZnS quantum dots) were mixed with raspberry particles. The concentration of raspberry particles was kept constant at 0.3 nM in all mixtures, while the nanoparticle concentrations in the mixtures were 0, 50, 250, 500, or 1250 nM. Films of these mixtures were prepared on a glass microscope slide, which was cleaned by rinsing it with isopropanol, acetone, and water, by drop-casting 500 μL on an area of ca. 2.5 cm x 3.6 cm.

*Structural characterizations of the nanoparticles.* High angle annular dark field scanning transmission electron microscopy (HAADF STEM) images were taken using a FEI Tecnai Osiris 200 kV transmission electron microscope. Specimens were prepared by drop-casting from solution on a graphene oxide on lacey carbon Cu grid by EM Resolutions. Scanning electron microscopy (SEM) images were taken with a Zeiss SIGMA VP field emission SEM and an FEI Nova field emission SEM. To take the cross-section images, samples were first drop casted on an aluminium foil, then the dry film was cut and Au/Pd was sputtered onto the film prior to imaging using an Emitech K575X sputterer at a current of 50 mA for 15 s. The estimated metal thickness was ca. 3.8 nm.

*Optical characterizations.* Absorbance measurements of the QD and transmittance measurements of the films were taken using a Cary 100 UV-Vis spectrophotometer. An uncoated microscope slide was used to measure the baseline. At least three transmittance measurements were taken at different locations of the films to average-out effects of the surface morphology differences on the solid films. The reported transmission spectra in the manuscript are the averages of these measurements unless otherwise stated. Reflectance measurements were recorded using an Olympus BX60 microscope equipped with a Thorlabs CCS 100 spectrometer and a halogen light source by using a standard reflection sample as reference. Optical microscope images were taken using the same microscope equipped with a Zeiss AxioCam MRc5 camera.

*Electromagnetic simulations.* Electromagnetic simulations were carried out using a commercially available finite-different time-difference simulator (Lumerical) in two dimensions. The scattering field was calculated using total-field & scattered-field light source[27] and the angular distribution of the scattered field in the far field was calculated using a frequency domain field and power monitor. The RBs were constructed using a polystyrene sphere with a dimeter of 200 nm together with surrounding polyacrylate hemispheres of 45 nm diameter on the surface of polystyrene sphere. 45 nm-sized polystyrene particles were randomly positioned in a 1 μm by 1 μm region while the boundaries of the simulation area were set as perfectly matched layers (PML). The density of polystyrene particles was changed between 100 and 2000 particles per μm$^2$.

**Results and Discussion**

Our raspberry particles are composed of a 200 nm-sized polystyrene particle surrounded by 45 nm-sized polyacrylate hemispheres. They were synthesized using a one-pot reaction technique in which we crosslinked acrylate and styrene using divinylbenzene in a water-ethanol solution (Figure 1a).[22] The high refractive index variation between the RB material and water is responsible for the strong scattering, giving their dispersions milky appearance even when they are dilute (Figure 1b). Interestingly, a red reflection band emerges at certain angles in dry, solid films due Bragg reflections when the raspberries are densely packed[22] while appearing white at other angles (Figure 1c and 1d).

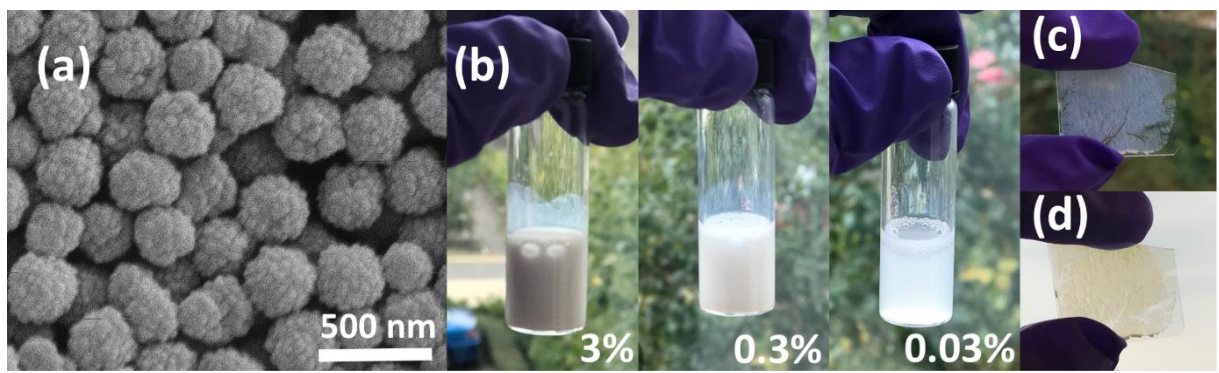

**Figure 1.** (a) SEM image of raspberry particles. (b) Photographs of raspberry particle aqueous dispersions at concentrations of 3 wt%, 0.3 wt% and 0.03 wt%. (c) and (d) Photographs of raspberry particle solid film at different angles.

To suppress this strong scattering in solid films, we hybridized our positively charged RBs with nanoparticles having various physical characteristics. First, we employed negatively-charged 45 nm-sized polystyrene (PS) nanoparticles, which are similarly sized as the bumps on the raspberries. To study the effect of the added nanoparticle amount, we varied the ratio of the polystyrene nanoparticles to raspberry particles as shown in Table 1. The final concentration of the raspberry particles in the final solution became 0.3 wt%, which we estimate to correspond to ca. 0.3 nM while the estimated molarity of the small polystyrene particles varied between 50 to 1250 nM. After mixing both types of particles at certain ratios, we left them to interact for at least 2 h and then drop-cast the suspension onto a microscope slide. The final surface area of the prepared films was ~9 cm$^2$ with thicknesses around 5-10 μm.

**Table 1**. Concentrations of the mixtures containing raspberry particles and negatively charged polystyrene particles along with the number of polystyrene particles per raspberry particle and the ratio of raspberry particle and polystyrene particles of different samples.

| Sample No. | RB Particle (nM) | PS Particle (nM) | Number of PS Particle per RB Particle | Ratio of PS Particle Volume to RB Particle Volume |
|---|---|---|---|---|
| 1 | 0.3 | - | 0 | 0 |
| 2 | 0.3 | 50 | 173 | 1.26 |
| 3 | 0.3 | 250 | 865 | 6.28 |
| 4 | 0.3 | 500 | 1730 | 12.6 |
| 5 | 0.3 | 1250 | 4325 | 31.4 |
| 6 | - | 250 | N/A | N/A |

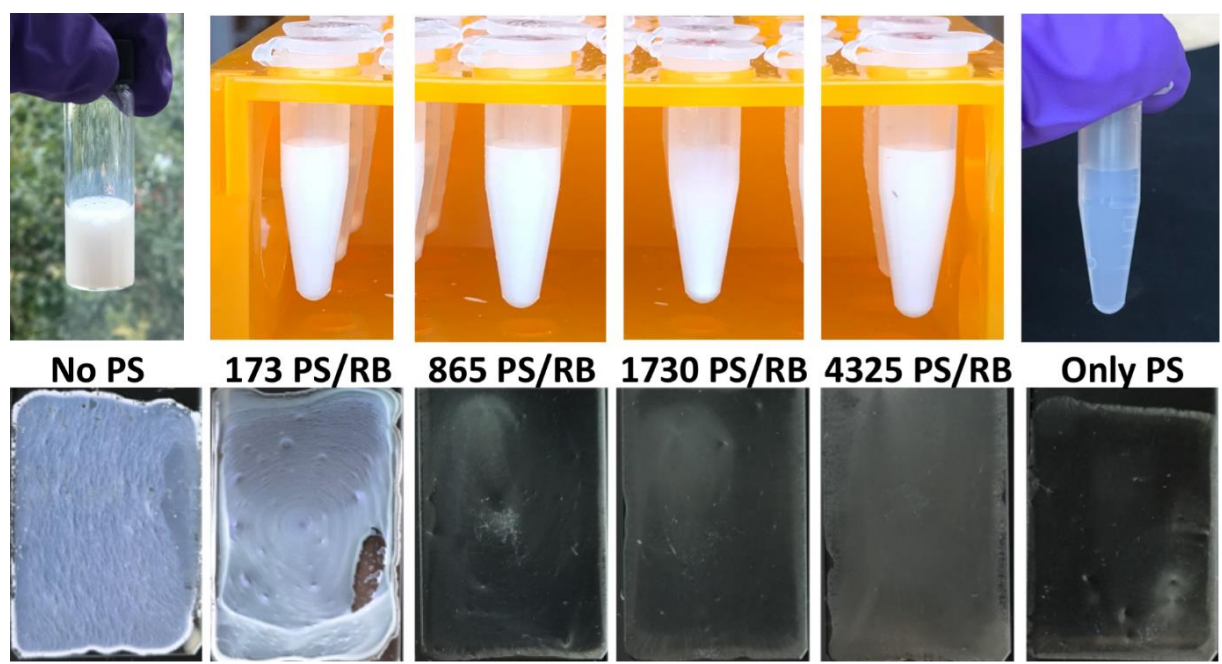

**Figure 2**. Photographs of pure raspberry particle (RB) suspensions (left; upper row) and the hybrid samples with varying negatively-charged polystyrene (PS)-to-RB particle ratios and only negatively-charged PS-particle dispersions (right; upper row) and the respective drop-cast solid films (lower row). All dispersions had a starting solid content of about 2 v%.

As shown in Figure 2, the film prepared without negatively-charged PS-nanoparticles exhibits strong scattering leading to a white appearance. However, increasing the ratio of PS-nanoparticles to raspberry (RB) particles surprisingly induces increased transparency of the films. Our measurements show that the film made of only raspberry particles has an average transmission above 70% at longer wavelengths whereas this transmission quickly decreases down to 50% at 550 nm and to 15% at 400 nm (Figure 3). We observe that the addition of oppositely-charged PS-particles to RB particles first affects the transmission in the blue-ultraviolet regime. For example, the sample prepared using a ratio of 173 PS particles to RB particles shows almost 30% transmission at 400 nm, while the transmission at 550 nm and 800 nm remains almost the same as that in pure RB-particle films. A further increase of the negatively-charged PS-to RB ratio to 865 dramatically boosted the transparency of the film to above 90% between 420 and 800 nm. Interestingly, further increasing in the PS-to-RB ratio to 1730 caused a slight decrease in the transmission of the film, while a more profound increase in the PS-particle number further pushes down the transmittance levels. That decrease was accompanied by the appearance of a resonant behavior in the transmission spectra. Since the data presented in Figure 3 are averaged over the measurements taken at different parts of the film, the resonant behavior seemed to be averaged out. These resonances can be seen more clearly in individual transmission spectra presented in Figure S1. A clear periodicity is also apparent when the transmission spectra are drawn against the photon energy rather than the wavelength (Fig. S1).

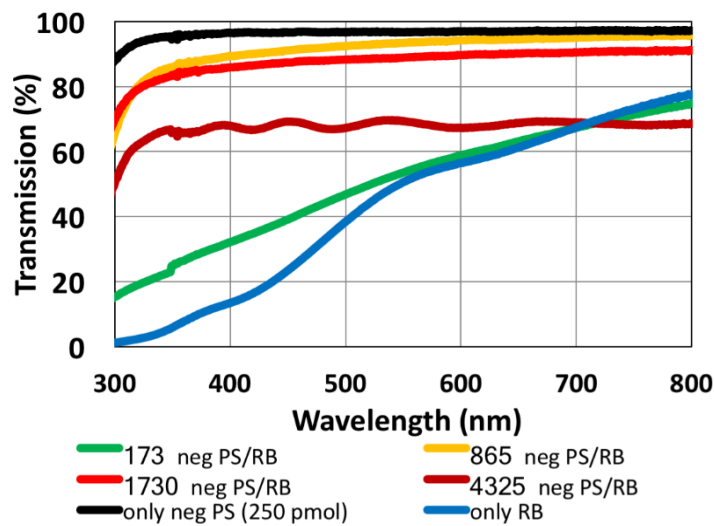

**Figure 3.** Averaged transmission spectra of the solid-films employing RB and negatively-charged PS particles (I suggest you use "-PS NP" for negatively charged PS nanoparticles and maybe +RB for positively charged raspberry particles throughout the paper.) at various particle number ratios.

Since the surface of the films may give insight into this resonant behavior, we imaged the surface of the films using optical microscopy (Figure 4). We found that the films composed of only raspberry particles have very rough surfaces with a brown-reddish appearance supporting our previous work[22]. In the sample prepared using a ratio of 865 PS nanoparticles to RB particle (corresponding to 6.26 volume ratio of particles, see Table 1), the surface is clearly flatter compared to the film prepared solely from RB particles and strongly resembles the film made of only PS particles. Further increasing the PS nanoparticle content interestingly causes the occurrence of cracking patterns that do not develop in the films made of only the small PS nanoparticles. These cracks become denser and more regular when more polystyrene particles are employed and resemble an optical grating causing the observed resonance behavior in transmission. The effect of these cracks is also apparent in the reflection spectra recorded using an optical microscope (Fig. S2). The appearance of cracks in drying colloidal films is well understood. In particular, the width between the typical long crack lines can be rationalized by the capillary-pressure build up in the drying films. A detailed discussion of this pattern formation is given in references [28-29]. At the locations of the film where these cracks are denser, we observe more resonance peaks accompanied by increased reflection. This behavior also indicates that these cracks are responsible for the decreased transmission and resonant behavior of the hybrid films, which employ higher PS-nanoparticle loadings.

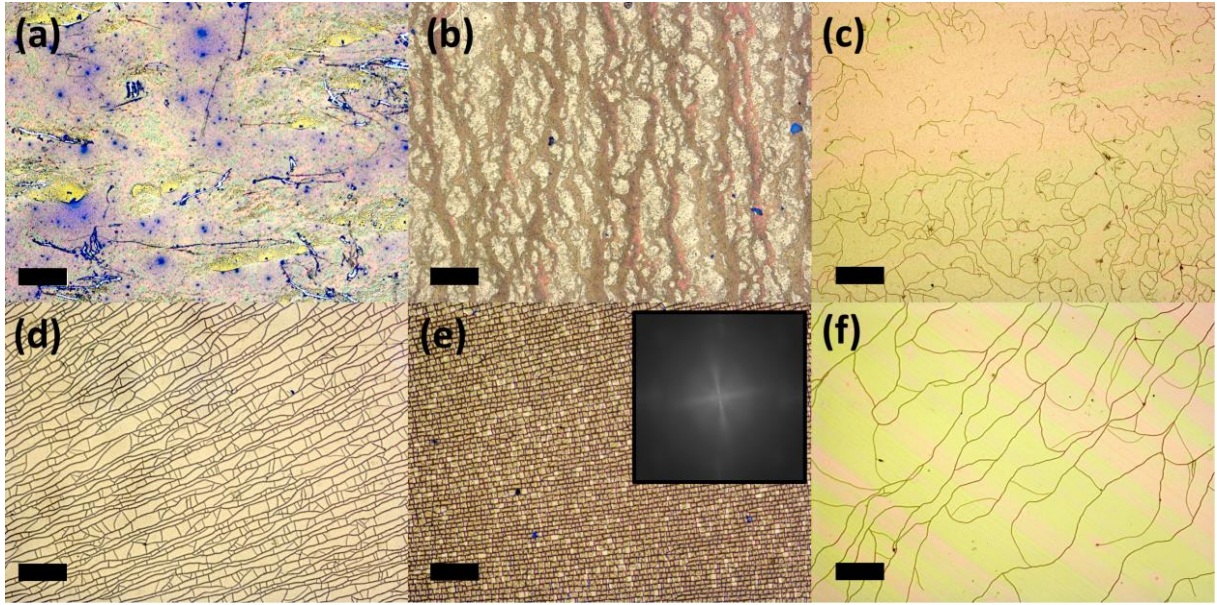

**Figure 4.** Optical microscopy images of the dried films made of (a) only raspberry particles along with hybrid samples prepared using (b) 173, (c) 865, (d) 1730 and (e) 4325 negatively charged polystyrene particles per raspberry particles with the fast Fourier transform of the image (inset), and (f) dry film of the negatively charged polystyrene particle. Scale bar: 200 µm.

To reveal the reason behind the observed transparency, we investigated the structure of the polystyrene and raspberry particle hybrid films using scanning electron microscopy (SEM). SEM images shown in Figure 5 indicate that some of the negatively-charged polystyrene nanoparticles surround the oppositely-charged raspberry particles arising from electrostatic interactions. More importantly, these new hybrid structures tend to be placed within a matrix of small polystyrene nanoparticles. Increasing transmission with increasing PS particles to RB ratios shows that these raspberry particles have a limited optical interaction with each other leading to transmission spectra similar to the film prepared from 45 nm PS particles alone.

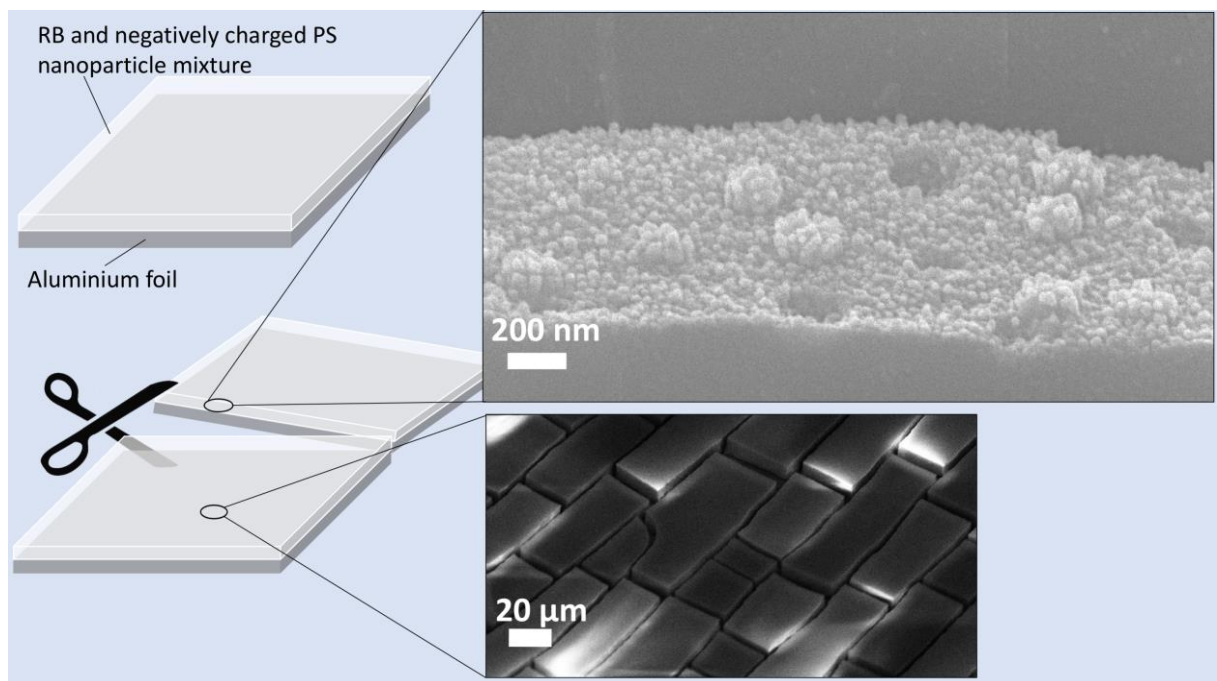

**Figure 5.** Illustration of the scanning electron microscope (SEM) sample preparation, cross-sectional SEM image of the raspberry particles and negatively charged polystyrene particle film of the sample prepared using 865 PS-to-RB-particle ratio (top, right) and SEM image of the film surface (bottom, right).

We also investigated dried films made purely of RB particles coated with the negatively-charged 45 nm PS particles. For this we centrifuged a PS-RB mixture with 1250 folds excess PS particles at low speed (1000 rpm), removed the supernatant containing the small PS particles and then redispersed the precipitate containing RB particles surrounded by PS particles in deionized water. After repeating this procedure five times, we prepared a film of the PS-coated RB particles solution by drop-casting onto a glass substrate. As presented in Fig. S3, the transparency of the film has slightly improved but remained much worse compared to the case without centrifugation. This result showed that the large number of smaller nanoparticles forming a matrix for the raspberry particles plays a crucial role in suppressing the scattering rather than having the small PS particles only on the RB particle.

We also studied the effect of the PS-particles on the scattering features of the RB-particles using finite-difference time-domain (FDTD) electromagnetic simulations. As the far field scattering spectra indicate (Fig. S4), RB-particles scatter the incident plane wave at almost all angles. On the other hand, addition of small polystyrene nanoparticles dramatically decreased the angular distribution of the scattering angles. This behavior supports our observation of suppressed scattering in hybrid films. Furthermore, the simulation results showed that the short wavelength part of the scattering spectrum also concentrates around zero degrees with increasing density of the smaller PS particles hybridized to RB particles, which is also observed in our experiments. Further, we investigated the far field scattering spectra of the small PS particles without RBs and found that the recorded spectra were similar to those of the hybrid structures. Thus, we conclude that the small polystyrene particles constituting the matrix are mainly responsible for suppressing the scattering as they avoid aggregation of RB particles and decrease the spatial variation of the refractive index.

To test the effect of the surface charge on the optical properties of the films, we mixed our positively-charged RB particles with positively-charged 45 nm PS particles. As opposed to the films with oppositely-charged PS particles, these films did not become transparent as shown in Figure S5. Their transmission spectra (Fig. S6) show no profound improvement; nevertheless, the reflection spectra measured with an optical microscope revealed an increased reflectance with increasing positively-charged PS particle content (Fig. S7). The optical microscopy images shown in Fig. S8 indicates the formation of the film with irregular dense cracks when a small number of polystyrene particles are employed. Increasing the ratio of polystyrene particles to raspberry particles triggers the formation of regular crack patterns on the film surface, which increases the reflection and causes its resonant behavior, similar to the films where negatively-charged PS particles were used. However, in this case the distribution of the raspberry and positively PS particles seem to be significantly different than the films prepared using RB and negatively charged PS particles (Fig. S9).

One important aspect for the understanding of the transparency in our composite films prepared from complementary electrostatic PS and raspberry particles is the effect of the refractive index. Hence, we also hybridized our raspberry particles with oppositely charged semiconductor InP/ZnS core/shell quantum dots with a size of ca. 10 nm. The dark field scanning transmission electron microscopy images presented in Figures 6a and S10 indicate that the quantum dots successfully surround the surface of the raspberry particles, similar to the films prepared using negatively-charged polystyrene particles.

Additionally, the polymer stabilizing the QDs enveloped the raspberries to the point that the optical features of these films started to resemble those made of the polystyrene particles (compare spectra in Figure 3 and 6). However, different to the films with polystyrene particles, we observed a slight decrease in the average transmission of the film prepared using a ratio of 173 quantum dots per raspberry particle compared with the film prepared from only raspberry particles alone. Further addition of quantum dots triggered a strong increase in the transparency of the films especially in the long wavelength regime while the absorption of the quantum dots does not allow the transmission to increase significantly in the short wavelength regime.

In contrast to negatively-charged PS-RB particle hybrids, here we did not see any resonant behavior in the transmission spectra. The optical microscopy images reveal that the surface roughness of the hybrid films remains high when the amount of quantum dots is low (Figure S11). Further increasing the amount of quantum dots per RB particle does not allow the raspberry particles to aggregate and causes a smooth film formation as also shown by SEM in Figure 6. This smoothness then translates into suppression of scattering arising from the raspberry particles especially at longer wavelengths. Moreover, no cracks developed in these films, which contributed to the increased transparency with increasing amounts of hybridized quantum dots. Finally, the absence of the cracks also led to transmission spectra without resonance peaks in contrast to the films prepared from the negatively-charged small PS particles.

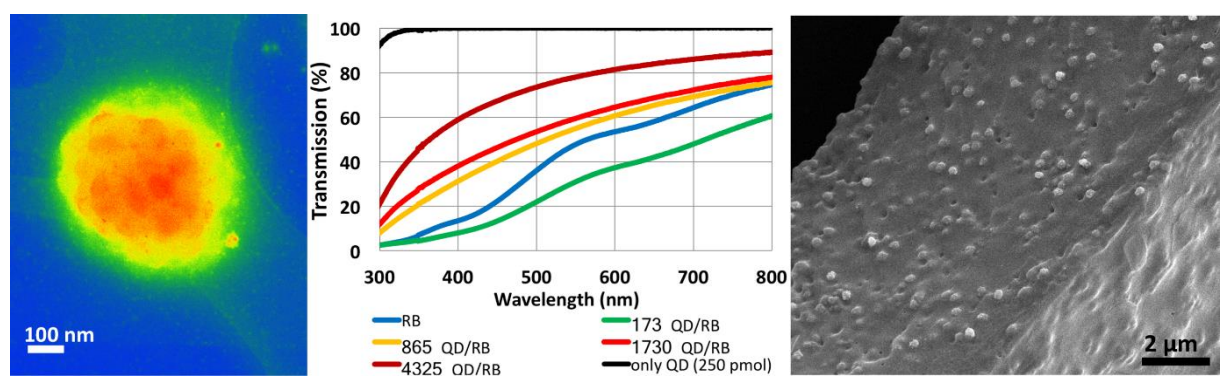

**Figure 6.** (Left) Coloured HAADF STEM image of raspberry particle (RB) and quantum dot (QD) hybrid (4325 quantum dot per raspberry particle sample). (Middle) Transmission spectra of the films made of quantum dot and raspberry particles hybridized at different number ratios.

(Right) SEM image of raspberry particle and quantum dot hybrid (4325 quantum dot per raspberry particle sample).

To understand whether the large surface area and the roughness of the raspberry particles plays a role on the observed optical properties, we carried out a control experiment employing 280 nm-sized, positively-charged, smooth polystyrene spheres instead of RBs. We hybridized these large particles with 45 nm-sized oppositely-charged polystyrene particles and formed their films. The photos and optical microscope images of the films presented in Figure S12 and S13 show that smaller negatively-charged PS particles have a profound effect on the transmission of the films with these large polystyrene particles, as was the case with RBs. Furthermore, the transmission measurements reveal that upon addition of the small nanoparticles, the transmission in short wavelengths increased significantly, which is similar to raspberry and negatively-charged PS particle hybrid films (Figure S14). Finally, the transmission levels and the trends of the transmission were similar to the films made of raspberry particles alone. These results also support our hypothesis that that small oppositely-charged nanoparticles filling the bumps of the RB particles act as a matrix owing to their high number and are mainly responsible for the observed variations in transmission while the raspberry particles themselves do not have a significant role on the observed phenomena.

**Conclusions**

In this work, we developed a simple strategy to obtain transparent films of highly scattering particles. With this motivation, we studied raspberry particles as a good example of strongly scattering nanoparticles. We found that solid films prepared from these scattering particles can be made transparent upon hybridization with oppositely-charged smaller nanoparticles, which can also form transparent films alone. Our experiments revealed that oppositely-charged smaller nanoparticles surround the surface of the larger, strongly scattering particles on account of electrostatic interactions. When the concentration of the smaller nanoparticles is high enough, these small particles also act as a matrix. These two effects aid in the lack of aggregation of scattering particles; as a result, a smooth film can be formed and the angular distribution of the scattering is strongly concentrated around the incidence angle leading to an observed transparency. Our experiments with smaller nanoparticles having the same charge as the scattering particles indicated that it was critical to use oppositely-charged particle systems to induce transparency as aggregation could be avoided. Furthermore, our experiments with quantum dots brought us to the conclusion that, if the particles have opposite charges, and the small nanoparticles possess transparency at least in some part of the spectrum, the transparency is induced regardless of the refractive index of the small nanoparticles. Finally, we revealed that the observed transparency upon hybridization does not depend on the shape of the large particles based on our experiments with large spherical and raspberry particles. We believe that the simple methodology of inducing transparency explored here may prove useful for various types of scattering organic and inorganic nanoparticles in optical coatings or optoelectronic devices in the future.

## Acknowledgements

TE acknowledges Royal Society for the Newton International Fellowship. We also thank Peian Li for optimizing the polystyrene nanoparticle syntheses.

# Inducing transparency in the films of highly scattering particles

Talha Erdem[1,*], Lan Yang[2], Peicheng Xu[1], Yemliha Altintas[3], Thomas O'Neil[1], Alessio Caciagli[1], Caterina Ducati[4], Evren Mutlugun[3], Oren A. Scherman[2], Erika Eiser[1,*]

[1]*Cavendish Laboratory, Department of Physics, University of Cambridge, JJ Thomson Avenue, Cambridge CB3 0HE, United Kingdom*

[2]*Melville Laboratory for Polymer Synthesis, Department of Chemistry, University of Cambridge, Lensfield Road, Cambridge CB2 1EW, United Kingdom*

[3]*Abdullah Gül University, Departments of Electrical-Electronics Engineering, and Materials Science and Nanotechnology Engineering, 38080, Kayseri, Turkey*

[4]*Department of Material Science and Metallurgy, University of Cambridge, 27 Charles Babbage Road, Cambridge CB3 0FS, United Kingdom*

*Corresponding Author: te294@cam.ac.uk and ee247@cam.ac.uk

## Supporting Information

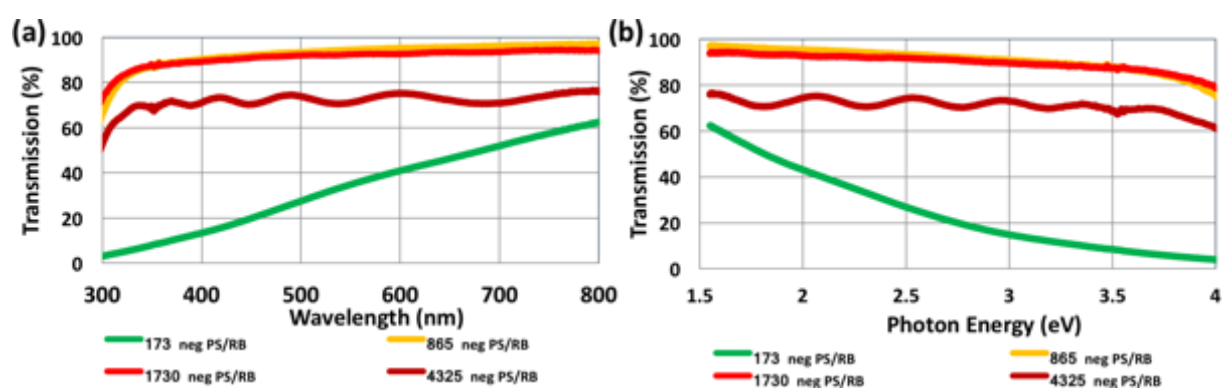

Figure S1. Transmission spectra of the solid-films employing raspberry particles (RB) and negatively charged polystyrene particles (neg PS) at various particle number ratios presented against (a) the wavelength and (b) photon energy. These measurements belong to individual measurements.

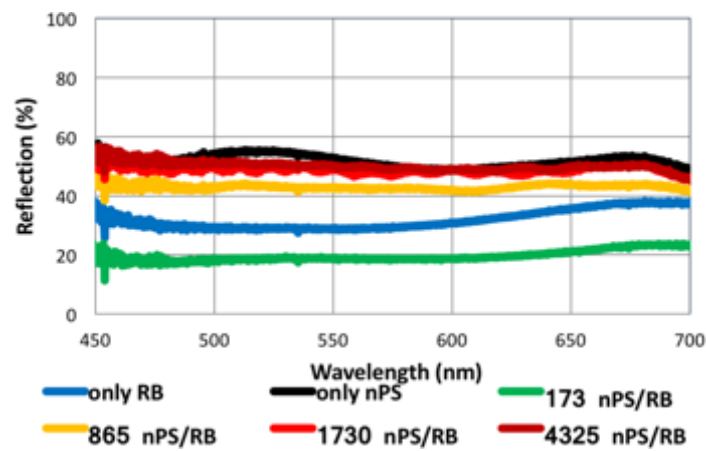

Figure S2. Averaged reflection spectra of the solid-films employing raspberry particles (RB) and negatively charged polystyrene particles (nPS) at various particle number ratios.

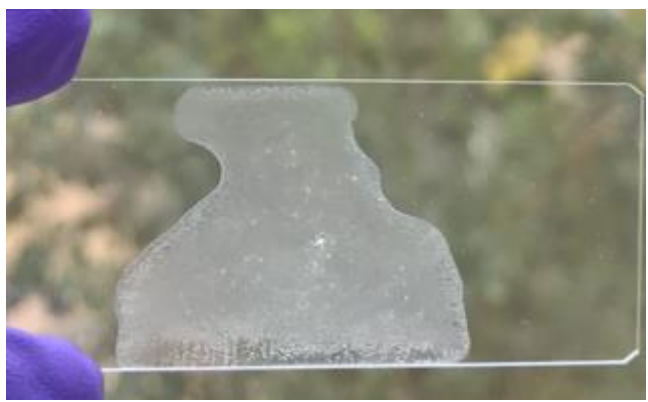

Figure S3. Solid film prepared after centrifuging the raspberry particle and negatively charged polystyrene particle mixture for five times to remove the polystyrene particles that are not attached to raspberry particles as much as possible.

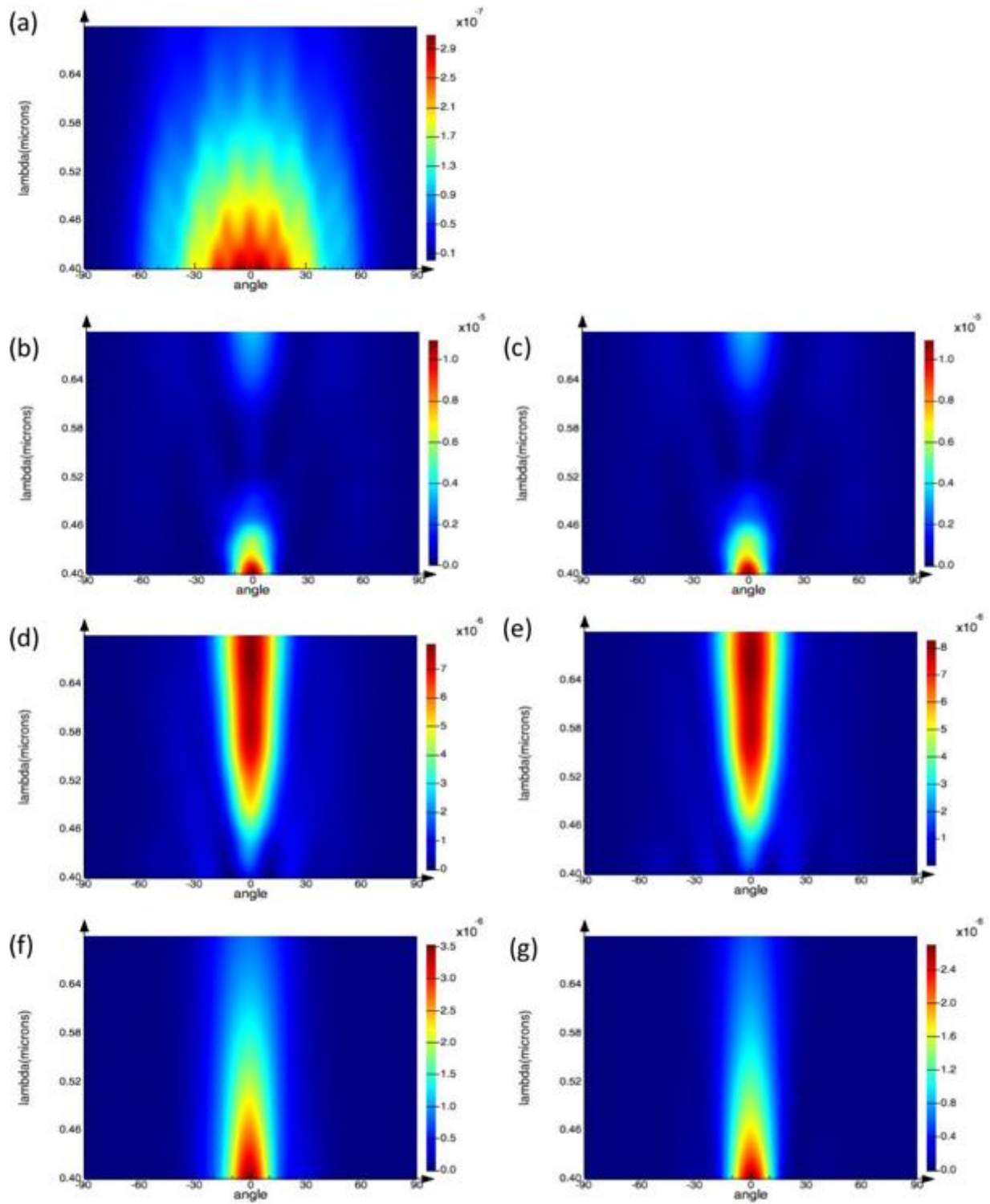

Figure S4. Scattering spectra calculated using two-dimensional finite-difference time-domain electromagnetic simulations. (a) Only raspberry particle, (b) raspberry particle and randomly distributed polystyrene particles (density: 100 particles/μm²), (c) only randomly distributed polystyrene particles (density: 100 particles/μm²), (d) raspberry particle and randomly

distributed polystyrene particles (density: 750 particles/μm²), (e) only randomly distributed polystyrene particles (density: 750 particles/μm²), (f) raspberry particle and randomly distributed polystyrene particles (density: 2000 particles/μm²), (g) only randomly distributed polystyrene particles (density: 2000 particles/μm²).

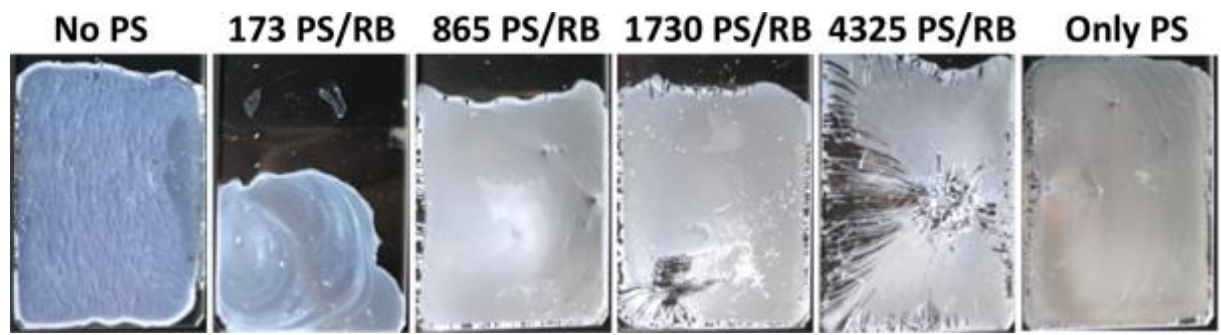

Figure S5. Photographs of the solid films made of only raspberry particles (no positively charged raspberry particle) sample together with hybrid samples with varying positively charged polystyrene-to-raspberry particle ratios and only positively charged polystyrene particles.

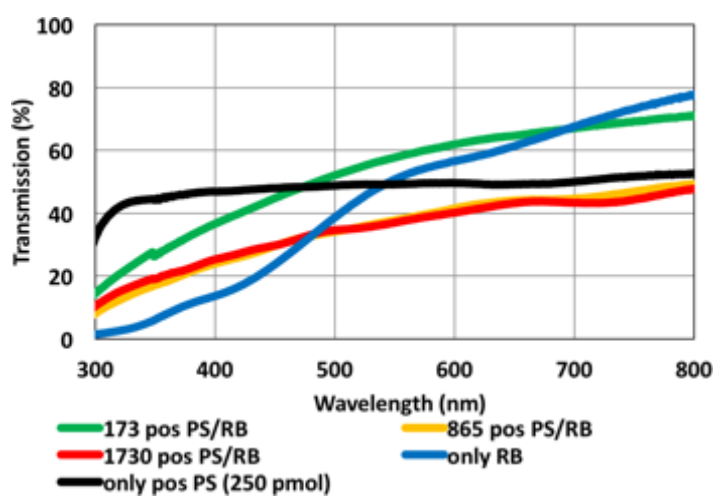

Figure S6. Averaged transmission spectra of the solid-films employing raspberry particles (RB) and positively charged polystyrene particles (pos PS) at various particle number ratios.

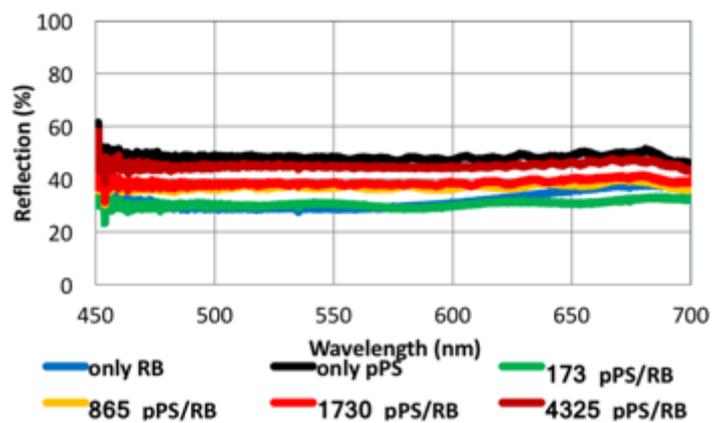

Figure S7. Averaged reflection spectra of the solid-films employing raspberry particles (RB) and positively charged polystyrene particles (pPS) at various particle number ratios.

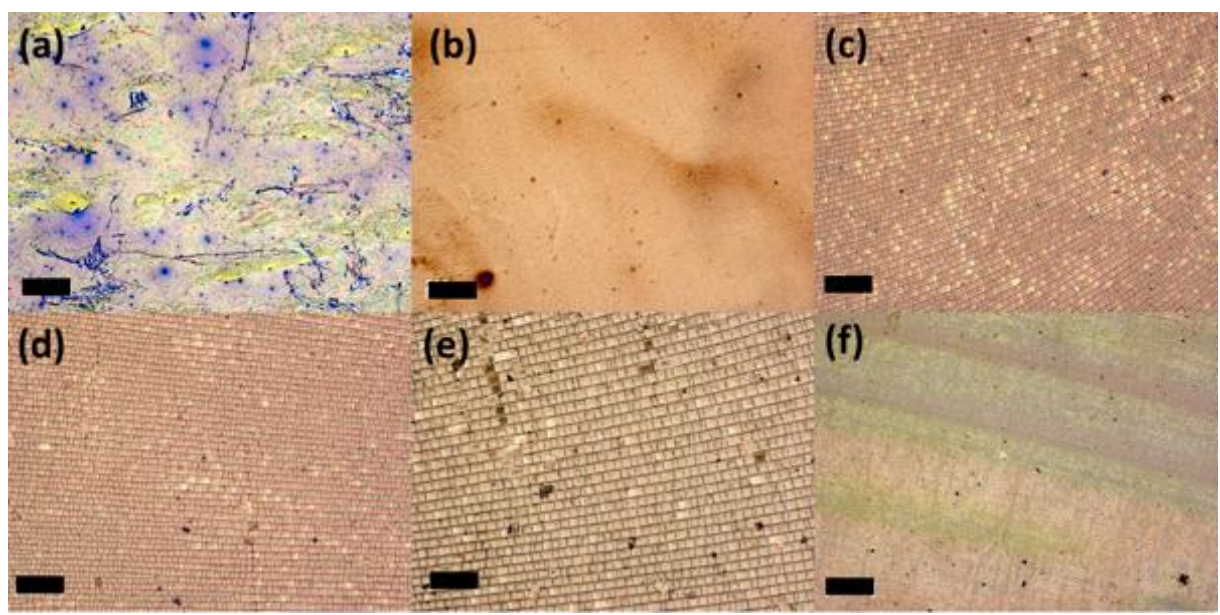

Figure S8. Optical microscopy images of the films made of (a) only raspberry particles along with hybrid samples prepared using (b) 173, (c) 865, (d) 1730 and (e) 4325 positively charged polystyrene particles per raspberry particles, and (f) only polystyrene particle film. Scale bar: 200 µm.

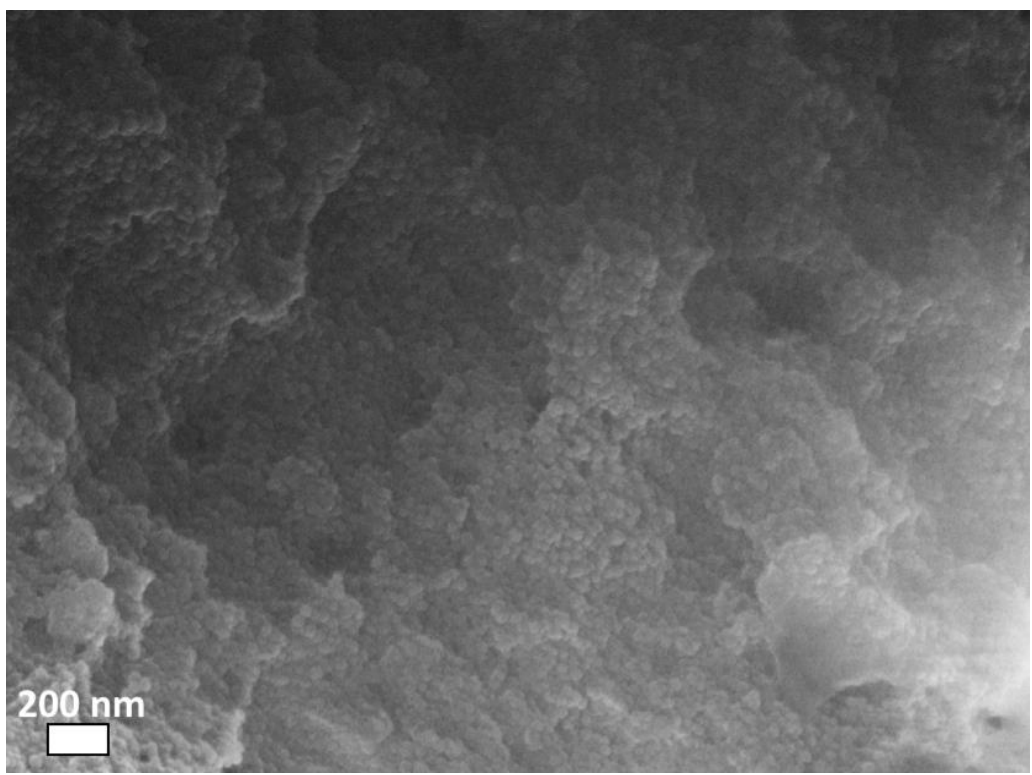

Figure S9. SEM image of the film made of RB particle and positively charged PS particles.

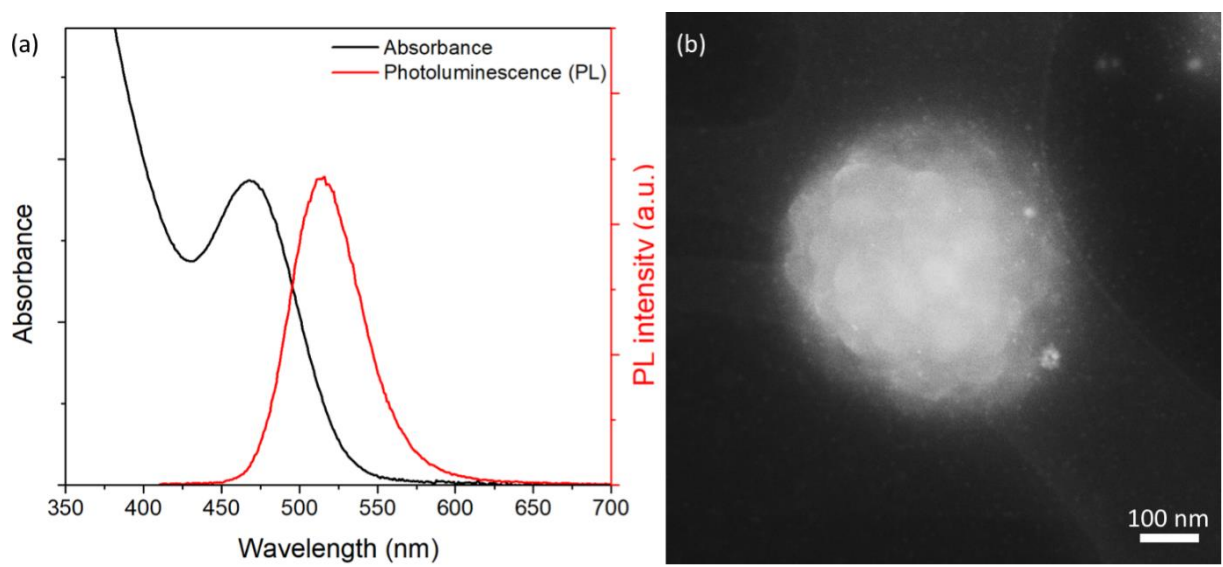

Figure S10. (a) Photoluminescence and absorbance spectra of the quantum dots. (b) HAADF STEM image of raspberry particle and quantum dot hybrid (4325 quantum dot per raspberry particle sample).

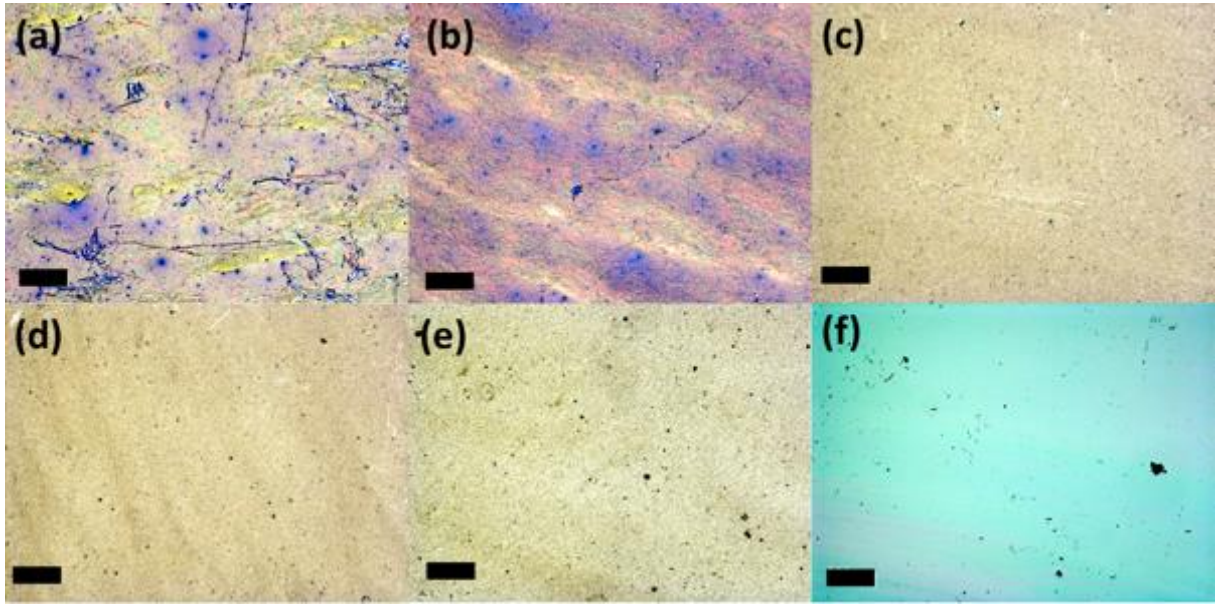

Figure S11. Optical microscopy images of the films made of (a) only raspberry particles along with hybrid samples prepared using (b) 173, (c) 865, (d) 1730 and (e) 4325 quantum dots per raspberry particles, and (f) only quantum dot film. Scale bar: 200 µm.

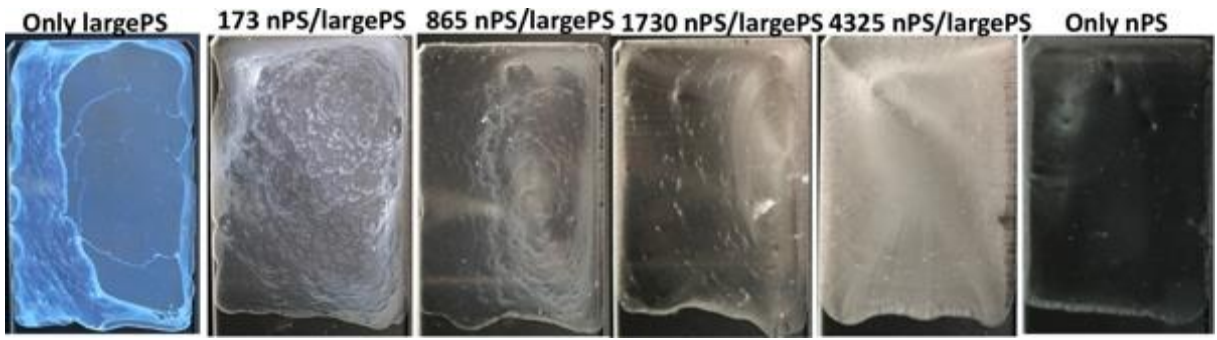

Figure S12. Photographs of the solid films made of only large positively charged polystyrene particles sample together with hybrid samples with varying large positively charged and small negatively charged polystyrene particle ratios and only negatively charged small polystyrene particles.

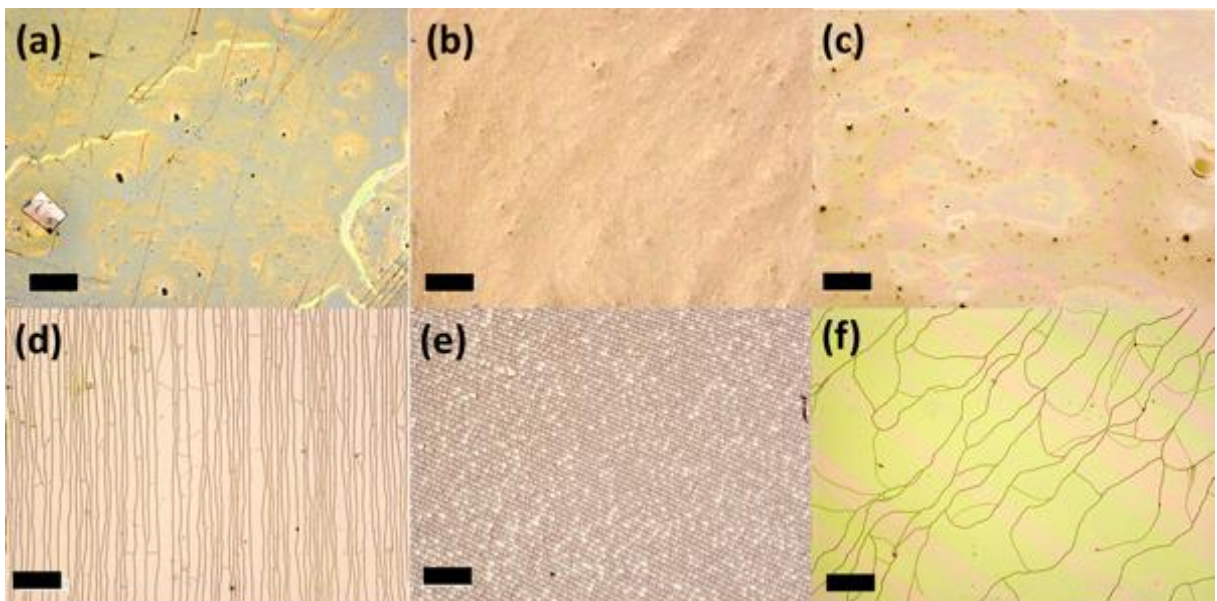

Figure S13. Optical microscopy images of the films made of (a) only large positively charged polystyrene particles along with hybrid samples prepared using (b) 173, (c) 865, (d) 1730 and (e) 4325 negatively charged small polystyrene per positively charged large particles, and (f) only negatively charged polystyrene particles. Scale bar: 200 µm.

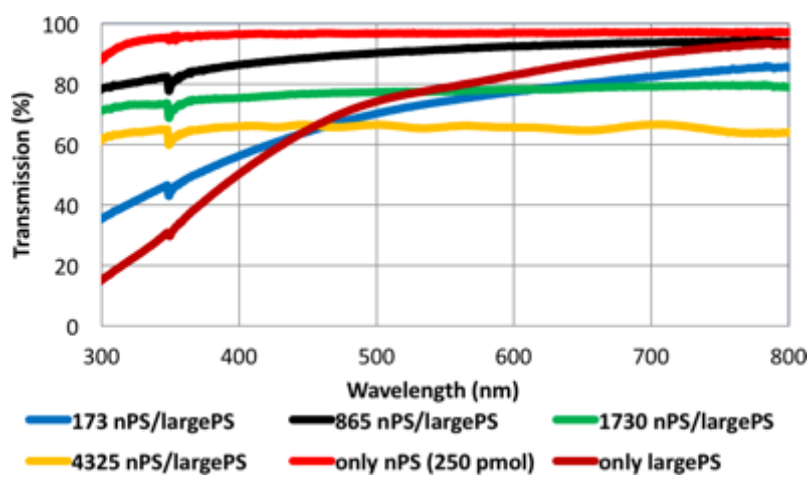

Figure S14. Averaged transmission spectra of the solid-films employing positively charged large polystyrene particles (largePS) and negatively charged polystyrene particles (nPS) at various particle number ratios.